# Extending ballistic graphene FET lumped element models to diffusive devices

**G Vincenzi[1,2], G Deligeorgis[1,2], F Coccetti[1,2], M Dragoman[3], L Pierantoni[4], D Mencarelli[4] and R Plana[1,2]**

[1] CNRS; LAAS; 7 avenue du Colonel Roche, F-31077 Toulouse Cedex 4, France

[2] Université de Toulouse; UPS, INSA, INP, ISAE; UT1, UTM, LAAS; F-31077 Toulouse Cedex 4, France

[3] National Institute for Research and Development in Microtechnology (IMT); P.O.Box 38-160, 023573 Bucharest, Romania

[4] Dipartimento di Ingegneria dell'Informazione; Università Politecnica delle Marche; I-60131 Ancona, Italy

## Abstract

In this work, a modified, lumped element graphene field effect device model is presented. The model is based on the "Top-of-the-barrier" approach which is usually valid only for ballistic graphene nanotransistors. Proper modifications are introduced to extend the model's validity so that it accurately describes both ballistic and diffusive graphene devices. The model is compared to data already presented in the literature. It is shown that a good agreement is obtained for both nano-sized and large area graphene based channels. Accurate prediction of Drain current and transconductance for both cases is obtained.

## 1. Introduction

Graphene's electronic band structure has been the subject of theoretical investigation more than 60 years ago [1-4]. However, the isolation of graphene in 2004 [5] allowed measuring its properties and verified the expected high electron mobility and reduced scattering [6]. Quantum models [4],[7] were readily accessible to the theoretical modeling of these properties at the time. These were also used for the treatment of Graphene Nanoribbon (GNR) devices [8-10]. GNRs are graphene stripes with a width in the nanometer scale. In field effect transistors based on such systems (GNRFETs), the drain current has been computed in the ballistic limit through Tight Binding (TB) models, solved using Non-Equilibrium Green's Functions (NEGF) formalism [11-13] or the scattering matrix approach [14], [15]. However, the severe computational load of such models – which scales with the dimensions of the device – limited their applicability to graphene device modeling. On the other hand, "Top-of-the-barrier" semiclassical models [16] have been applied to GNRFET for ballistic [11],[17] and tunneling/thermionic [18] carrier transport, bypassing the computational load limit. Unfortunately, the latter are valid only for short-channel devices which limits their use.

Most experiments on graphene conductivity properties were based on Graphene FET (GFET) devices with gate length ranging from hundreds of nm to several microns. Graphene of those dimensions is called large-area graphene because such systems are characterized by carrier diffusion [19],[20] rather than ballistic transport. Semiclassical models that are applied to GFET structures followed different approaches such as drift-diffusion semi-analytical models [21-24], the virtual-source semi-empirical model [25] and analytical models for both GFET [26] and GNRFET [27],[28]. The first two model types were validated using DC measurements of GFETs of different dimensions, but they failed to correctly predict the behavior of a ballistic GNR FET device. The last family instead has not been experimentally validated to our knowledge.

In this work a model that can correctly predict both ballistic as well as diffusive transport and thus describe all graphene field effect devices is presented and validated. It is based upon the "Top-of-the-barrier" model for ballistic graphene nanotransistors [16], [17], which is extended to correctly predict the behavior of long-channel diffusive graphene FETs. Section 2 presents the "Top-of-the-barrier" model and details the modifications necessary to extend its validity to large-area graphene. Section 3 presents model validation comparing modeling and experimental results for two FET devices selected from literature.

## 2. Theoretical model

"Top-of-the-barrier" models [16], [17] have been successfully used to model ballistic GFET's. The structure of a typical GNR FET is shown in Fig 1. According to this approach, the drain current is given by:

$$I_{\mathrm{d}} = I_{\mathrm{n}} - I_{\mathrm{p}} \,, \qquad (1)$$

where the $I_n$ and $I_p$ correspond to electron and hole current respectively. A similar convention for all variables is followed throughout this work. In a ballistic nanotransistor, mobile carriers are injected directly from the contacts to the channel without undergoing an energy relaxation process into the channel's statistics [29]. The net current is given by the difference of the injected electron and hole currents. The electron current is given by the Landauer equation of conduction [29], [30]:

$$I_{\mathrm{n}} = \frac{\overline{v_{\mathrm{x}}} q}{2\hbar} \int_{-\infty}^{+\infty} D_{\mathrm{n}}(E)\,(f_{\mathrm{S}} - f_{\mathrm{D}})\,\mathrm{d}E \,. \qquad (2)$$

where $\overline{v_{\mathrm{X}}}$ is the carrier velocity, D(E) is the density of states and $f_S$ and $f_D$ are the Fermi functions for the two contacts. As already described, for ballistic transport, injected carriers are described by the Fermi potential of their respective contact. The velocity of carriers $\overline{v_{\mathrm{X}}}$, is calculated by the carrier dispersion relation of the channel material. In the case of a GNR, this is obtained as described in [17].

The hole current is computed in a similar fashion. Differentiating the dispersion relation and averaging over all available subbands and k-space yields the band velocity:

$$\overline{v_{\mathrm{x}}} = \frac{1}{2\pi} \int_{-\pi/a}^{+\pi/a} \frac{1}{\mathrm{N}} \sum_{n=1}^{\mathrm{N}} \left| \frac{1}{\hbar} \frac{\mathrm{d}\epsilon_n(k)}{\mathrm{d}k} \right| \mathrm{d}k \,, \tag{3}$$

where k is the wavenumber, $e_n(k)$ is the energy of the n-th sub-band of the material and N is the number of unit cells. In [16] the carrier velocity is expressed as function of energy; carriers are injected at the top of the barrier where their velocity is at their lowest and cannot accelerate along the channel. This underestimates their overall velocity in the Top of the barrier model. The use of an averaged value of $v_x$ as in Eq. (3) is then a more convenient choice for both accuracy and simplicity. The density of states D is analytically derived from the bandstructure [31]. To account for impurities and for the gate metal work function, the neutrality point is shifted by the Dirac Voltage not shown here. The Dirac voltage is a free parameter in the model [16]. Assuming the source contact potential is set as the reference, the drain potential is given by $-qV_{ds}$, where q is the electron charge and $V_{ds}$ is the drain-source voltage difference. For this case, the expressions of the Fermi Dirac statistic $f_S$ for the source and $f_D$ for the drain become:

$$\begin{aligned} f_{\mathrm{S}}(E) &\equiv f_{\mathrm{FD}}(E)\,, \\ f_{\mathrm{D}}(E) &\equiv f_{\mathrm{FD}}(E - qV_{\mathrm{ds}})\,, \end{aligned} \tag{4}$$

where $f_{FD}$ is the Fermi-Dirac statistics equation.

To take into account the electrostatics of the problem, all the previous quantities are shifted in energy by the electrostatic potential U. This latter is the sum of two terms [30], the charge-less potential $U_L$ and the mobile charge potential $U_P$. The expression of the total potential U is thus:

$$U = U_{\mathrm{L}} + U_{\mathrm{P}}\,. \tag{5}$$

The electrostatic part is computed using a lumped element model. A capacitor network consisting of a gate capacitance $C_G$, a source $C_S$ and drain $C_D$ capacitance [16], [30] are taken into account. For simplicity it is assumed that $C_S=C_D$. The source and drain capacitance is a fit parameter [16]. Taking into account that $V_S=0$, the expression of the Laplace potential $U_L$ becomes:

$$U_{\mathrm{L}} = -q\left(C_{\mathrm{G}}V_{\mathrm{G}} + C_{\mathrm{D}}V_{\mathrm{D}}\right)\,. \tag{6}$$

$U_P$ is calculated using the linearized Poisson equation, where the potential is proportional to the charge unbalance in the channel [29]. For ambipolar graphene it is:

$$U_{\mathrm{P}} = \frac{q^2}{C_\Sigma}\left(N - N_0 - P + P_0\right)\,, \tag{7}$$

where $C_\Sigma$ is the sum of the three capacitors described above, N and $N_0$ are the number of mobile and fixed electrons in the channel respectively [30], P and $P_0$ are the number of holes. The inter-dependence between potential and charge concentration is solved self-consistently. As described in [16], the mobile charge concentration is computed taking the average of the two Fermi statistics that correspond to the contacts as these are described by eq.(4).

$$N = \int_{-\infty}^{+\infty} D(E)\cdot\left(\frac{f_{\mathrm{S}} + f_{\mathrm{D}}}{2}\right)\,\mathrm{d}E\,. \tag{8}$$

Fig 2(a) depicts the channel under the injection of carriers from the source and the drain contact, the corresponding pseudo-Fermi levels are labeled “S” and “D” respectively. The source injects carriers – in this case holes – which are described by its pseudo-Fermi potential and move to the right; similarly, the drain injects carriers moving to the left.

Equations (5), (7) and (8) are iteratively computed. Finally equations (3) and (4) are used to compute the current for a given bias taking into account equations (2) and (1).

The model described so far, is unable to correctly account for carrier scattering. Thus it breaks down for large channel length where diffusion transport becomes dominant. In order to extend the validity of this model to large-area GFET, we propose the following modification.

Assuming the distribution of scatterers is uniform across the channel, the transmission probability T of ballistic carriers scales inversely with gate length [29]. When the gate length is small compared to the mean free path, i.e. T~1, carriers that originate from the source contact are all described by the source pseudo-Fermi potential. Respectively those that originate from the drain are all described from the drain pseudo-Fermi potential. As the gate length increases, the value of T declines towards 0. The phenomenon is better understood looking into the carriers that move in the channel. The portion of carriers that ballistically traverse the channel decrease in number, while it is assumed that those that are scattered regain the thermodynamic equilibrium. In this work the elastic and inelastic scattering processes are considered to be intimately related, avoiding any effort to conceptually divide those two processes. Each elastic scattering event is considered to be followed by thermalization, which is thought as inevitably present as assumed in [29]. Hence carriers that have scattered should be described by the channel Fermi level.

Fig 2(b) shows a simplified case in which the charge scattering process breaks the drain flux in two parts: a fraction T that is transmitted and a fraction (1-T) that is scattered. To restore the apparent reduction of carriers introduced into the ballistic model by using the T factor, a modified, effective pseudo-Fermi potential φ is defined. As described in [29], φ is the weighted average of the contact pseudo-Fermi and the channel Fermi level. The weight factor is merely the transmission probability T. This way, all carriers originating from the drain contact are described by the effective level φ. A parameter λ related to the mean free path is introduced to describe the transmission factor T. The relation between T and the length coordinate x [29] is given by:

$$T = \frac{\lambda}{\lambda + x} \tag{9}$$

In order to obtain a value for φ that can be used in a lumped model the average of T over the entire length of the channel is used:

$$\begin{aligned} \varphi &= \left[1 - \frac{1}{L}\int_0^L \frac{x}{L}\,(1-T)\,\mathrm{d}x\right] qV_{\mathrm{ds}} \\ &= (1-k)\,qV_{\mathrm{ds}}\,, \end{aligned} \tag{10}$$

The level φ depends linearly from the parameter λ , and scales inversely with the gate length.

This modification to the "Top-of-the-barrier" models allows simulating GFETs with gate lengths larger than the mean fee path. In the case of short channel compared to the mean free path, this model reduces to the "top of the barrier" representation.

## 3. Model Validation

The proposed model is validated against two devices presented in the literature. A narrow channel graphene nanoribbon transistor described in [12] and a large area wide channel graphene FET described in [21]. The two devices are shortly presented in Table 1.

**Table 1.** Structure dimensions

| | FET1 | FET2 |
|---|---|---|
| $L$ | 15 nm | 3000 nm |
| $W$ | 1.35 nm | 2100 nm |
| $t_{\mathrm{ox}}$ | 1 nm | 15 nm |
| $\epsilon_{\mathrm{r}}$ | 3.9 | 16 |
| $V_0$ | / | 2.45 V |
| $V_{\mathrm{bs}}$ | / | -40 V |

The first device simulated is a GNR nanotransistor [12] with gate length L=15 nm, based upon a semiconducting nanoribbon. For clarity, the device cross section is sketched in Fig.1. The parameters used by the model are calculated using the procedure described in Section 2. The optimum results are obtained for a λ value of 21 nm. The current flowing through the device for two distinct drain voltages is depicted in Fig 3.

The quality of the agreement to the simulated currents from [12] is excellent for the low drain bias curve, while it's sub-optimum for the high drain bias one. The cause of this discrepancy is that for ballistic devices the act of referring the source contact to the 0 eV potential throughout the transistor's dynamic range is a condition too strong. After [16], the solution to this issue is to empirically identify a reference potential for *each* bias point. This procedure has not been endorsed by the authors of this work, in order to contain the model complexity at the expense of its accuracy. Moreover, the limited presence of scattering in the simulation of this ballistic transistor has been proven to be not a concern to the model accuracy. Finally, table 2 shows an excellent agreement of simulated $g_m$ with that presented in [12].

TABLE II
SIMULATED TRANSCONDUCTANCE $g_{\mathrm{m}}$ FOR FET1

| $V_{\mathrm{ds}}$ | [12] | This model |
|---|---|---|
| 0.1 V | 3.6 mS$\mu$m$^{-1}$ | 3.55 mS$\mu$m$^{-1}$ |
| 0.5 V | 4.8 mS$\mu$m$^{-1}$ | 4.81 mS$\mu$m$^{-1}$ |

The second device is a large-area graphene FET [21] with gate length L=3 µm and width W=2.1 µm (see Table 1). The structure of the FET (shown in Figure 4) is slightly more complex because of the back-gate electrode. This can be taken into account by adding a backgate capacitance term in equation (7) which thus becomes:

$$U_{\mathrm{L}} = -q\left(C_{\mathrm{B}}V_{\mathrm{b}} + C_{\mathrm{G}}V_{\mathrm{g}} + C_{\mathrm{D}}V_{\mathrm{d}}\right) . \tag{11}$$

where $V_b$ is the potential of the back-gate electrode and $V_g$ is the potential of the top-gate electrode. The values of the capacitors $C_G$ and $C_B$ have been computed using a finite elements simulator. The use of large-area graphene also implies the use of the related density of states equation, as reported in [32].

The parameters for the large-area graphene case remain the same compared to the GNR model, only the procedure to extract them is different. First, the Dirac voltage is empirically found by matching the position of the minimum conductivity point in the $I_d(V_{gs})$ transfer characteristic. According to the discussion in reference [21], the gate voltage – which corresponds to minimum conduction – is $V_{gs}$=2.38V thus the computed Dirac voltage has a value of $V_{Dirac}$=0.213 V.

The λ parameter is calculated using the $I_d(V_{ds})$ for low $V_{DS}$. In effect, the λ parameter will have a pronounced effect on the onset of the saturation effects in the $I_d(V_{gs})$ curve. In this case the calculated optimum λ=380nm which is consistent with the value found in literature [32].

Finally, the value of the average carrier velocity is found empirically by matching the magnitude of the current. The best match was obtained at $v_{\mathrm{X}}$ =3.25e4 ms$^{-1}$. A direct comparison with literature is not possible, since drift-diffusion models use a saturation model, where the drift velocity is an empirical function of the longitudinal electric field [34]. The value of carrier velocity used in this model is however consistent with the drift velocity that may be computed in the literature for similar structures [21], [23].

The output current $I_d(V_{sd})$ is plotted in Fig.5. The points correspond to DC current measurement taken from ref [21], whereas the family of solid lines is the predicted values using this model. It should be pointed out that although the model is based on a lumped element approach, unlike models described in ref [21],[23] and [24], it is able to correctly predict the presence of saturation effects in the current – voltage characteristics of the device. To our knowledge the only lumped element model presented so far with ability to predict the saturation and second linear characteristics of GFET's was based on different equations for each region. Finally, Table3 presents the peak measured and simulated $g_m$ for two different gate biases. The agreement is not excellent but this could be attributed to the $g_m$ calculation method used in [21] to remove the effect of the contact on the device behavior.

TABLE III
PEAK TRANSCONDUCTANCE $g_{\mathrm{m}}$ FOR FET2

| $V_{\mathrm{gs}}$ | [21] | This model |
|---|---|---|
| -1.5 V | 211 $\mu$S | 180 $\mu$S |
| -2.9 V | 205 $\mu$S | 186 $\mu$S |

In summary, the described model uses four parameters to fit the experimental results. Those are presented in Table 4, for the two simulated devices, with the exception of the velocity of FET1 which is analytically derived using (3).

TABLE IV
FREE PARAMETERS FOR FET1 AND FET2

| Name | FET1 | FET2 |
|---|---|---|
| $\overline{v_{\mathrm{x}}}$ | Eq. (3) | $3.25 \times 10^4$ ms$^{-1}$ |
| $V_{\mathrm{Dirac}}$ | -0.062 V | 0.213 V |
| $C_{\mathrm{D}}$ | $4.41 \times 10^{-20}$F | $6.12 \times 10^{-17}$F |
| $\lambda$ | 21 nm | 380 nm |

## 4. Conclusions

A simple modification to the "top of the barrier" model that enables accurate simulation of a broad range of graphene based transistors was presented. The model retains the simplicity of a lumped element approach and is able to correctly describe the I-V characteristics of both ballistic and diffusive devices. Furthermore, it is able to correctly predict the behavior of both large-area as well as graphene nanoribbon based field effect devices. Finally the presented model achieves a new level of generic graphene device modeling with a simplicity that makes it easier to use compared to complex detailed models based on NEGF or Tight Binding thus it is a favorable solution for use in graphene enabled circuit simulation tools.

# References

[1] P. Wallace, “The Band Theory of Graphite,” Physical Review, vol. 71, no. 9, pp. 622-634, May 1947. \cite{Wallace1947}

[2] J. McClure, “Band Structure of Graphite and de Haas-van Alphen Effect,” Physical Review, vol. 108, no. 3, pp. 612-618, Nov. 1957. \cite{McClure1957}

[3] J. Slonczewski and P. Weiss, “Band Structure of Graphite,” Physical Review, vol. 109, no. 2, pp. 272-279, Jan. 1958. \cite{Slonczewski1958}

[4] A. Castro Neto, F. Guinea, N. Peres, K. Novoselov, and A. Geim, “The electronic properties of graphene,” Reviews of Modern Physics, vol. 81, no. 1, pp. 109-162, Jan. 2009. \cite{CastroNeto2009}

[5] K. S. Novoselov et al., “Electric field effect in atomically thin carbon films.,” Science (New York, N.Y.), vol. 306, no. 5696, pp. 666-9, Oct. 2004. \cite{Novoselov2004}

[6] K. S. Novoselov et al., “Two-dimensional gas of massless Dirac fermions in graphene.,” Nature, vol. 438, no. 7065, pp. 197-200, Nov. 2005. \cite{Novoselov2005}

[7] S. Reich, J. Maultzsch, C. Thomsen, and P. Ordejón, “Tight-binding description of graphene,” Physical Review B, vol. 66, no. 3, pp. 1-5, Jul. 2002. \cite{Reich2002}

[8] Y.-W. Son, M. L. Cohen, and S. G. Louie, “Energy Gaps in Graphene Nanoribbons,” Physical Review Letters, vol. 97, no. 21, p. 216803, Nov. 2006. \cite{Son2006}

[9] L. Yang, C.-H. Park, Y.-W. Son, M. Cohen, and S. Louie, “Quasiparticle Energies and Band Gaps in Graphene Nanoribbons,” Physical Review Letters, vol. 99, no. 18, pp. 6-9, Nov. 2007. \cite{Yang2007}

[10] E. Mucciolo, A. Castro Neto, and C. Lewenkopf, “Conductance quantization and transport gaps in disordered graphene nanoribbons,” Physical Review B, vol. 79, no. 7, p. 075407, Feb. 2009. \cite{Mucciolo2009}

[11] Y. Ouyang, Y. Yoon, J. K. Fodor, and J. Guo, “Comparison of performance limits for carbon nanoribbon and carbon nanotube transistors,” Applied Physics Letters, vol. 89, no. 20, p. 203107, 2006. \cite{Ouyang2006}

[12] G. Fiori and G. Iannaccone, “Simulation of Graphene Nanoribbon Field-Effect Transistors,” IEEE Electron Device Letters, vol. 28, no. 8, pp. 760-762, Aug. 2007. \cite{Fiori2007}

[13] Y. Ouyang, P. Campbell, and J. Guo, “Analysis of ballistic monolayer and bilayer graphene field-effect transistors,” Applied Physics Letters, vol. 92, no. 6, p. 063120, 2008. \cite{Ouyang2008}

[14] D. Mencarelli, T. Rozzi, and L. Pierantoni, “Coherent carrier transport and scattering by lattice defects in single- and multibranch carbon nanoribbons,” Physical Review B, vol. 77, no. 19, p. 195435, May 2008. \cite{Mencarelli2008}

[15] D. Mencarelli, T. Rozzi, and L. Pierantoni, “Scattering matrix approach to multichannel transport in many lead graphene nanoribbons.,” Nanotechnology, vol. 21, no. 15, p. 155701, Apr. 2010. \cite{Mencarelli2010}

[16] A. Rahman, S. Datta, and M. S. Lundstrom, “Theory of ballistic nanotransistors,” IEEE Transactions on Electron Devices, vol. 50, no. 9, pp. 1853-1864, Sep. 2003. \cite{Rahman2003}

[17] G. Liang, N. Neophytou, D. E. Nikonov, and M. S. Lundstrom, “Performance Projections for Ballistic Graphene Nanoribbon Field-Effect Transistors,” IEEE Transactions on Electron Devices, vol. 54, no. 4, pp. 677-682, Apr. 2007. \cite{Liang2007}

[18] D. Jiménez, “A current-voltage model for Schottky-barrier graphene-based transistors.,” Nanotechnology, vol. 19, no. 34, p. 345204, Aug. 2008. \cite{Jimenez2008}

[19] J.-H. Chen, C. Jang, S. Xiao, M. Ishigami, and M. S. Fuhrer, “Intrinsic and extrinsic performance limits of graphene devices on SiO2.,” Nature nanotechnology, vol. 3, no. 4, pp. 206-9, Apr. 2008. \cite{Chen2008a}

[20] V. E. Dorgan, M.-H. Bae, and E. Pop, “Mobility and saturation velocity in graphene on SiO[sub 2],” Applied Physics Letters, vol. 97, no. 8, p. 082112, 2010. \cite{Dorgan2010}

[21] I. Meric, M. Y. Han, A. F. Young, B. Ozyilmaz, P. Kim, and K. L. Shepard, “Current saturation in zero-bandgap, top-gated graphene field-effect transistors.,” Nature nanotechnology, vol. 3, no. 11, pp. 654-9, Nov. 2008. \cite{Meric2008}

[22] I. Meric, N. Baklitskaya, P. Kim, and K. L. Shepard, “RF performance of top-gated, zero-bandgap graphene field-effect transistors,” in IEEE International Electron Devices Meeting, 2008. \cite{Meric2008b}

[23] S. A. Thiele, J. A. Schaefer, and F. Schwierz, “Modeling of graphene metal-oxide-semiconductor field-effect transistors with gapless large-area graphene channels,” Journal of Applied Physics, vol. 107, no. 9, p. 094505, 2010. \cite{Thiele2010a}

[24] S. Thiele and F. Schwierz, “Modeling of the steady state characteristics of large-area graphene field-effect transistors,” Journal of Applied Physics, vol. 110, no. 3, p. 034506, 2011. \cite{Thiele2011}

[25] H. Wang, A. Hsu, J. Kong, D. A. Antoniadis, and T. Palacios, “Compact Virtual-Source Current–Voltage Model for Top- and Back-Gated Graphene Field-Effect Transistors,” IEEE Transactions on Electron Devices, vol. 58, no. 5, pp. 1523-1533, May 2011. \cite{Wang2011}

[26] V. Ryzhii, M. Ryzhii, A. Satou, T. Otsuji, and N. Kirova, “Device model for graphene bilayer field-effect transistor,” Journal of Applied Physics, vol. 105, no. 10, p. 104510, 2009. \cite{Ryzhii2009}

[27] M. Ryzhii, a Satou, V. Ryzhii, and T. Otsuji, “High-frequency properties of a graphene nanoribbon field-effect transistor,” Journal of Applied Physics, vol. 104, no. 11, p. 114505, 2008. \cite{Ryzhii2008a}

[28] V. Ryzhii, M. Ryzhii, a Satou, and T. Otsuji, “Current-voltage characteristics of a graphene-nanoribbon field-effect transistor,” Journal of Applied Physics, vol. 103, no. 9, p. 094510, 2008. \cite{Ryzhii2008b}

[29] S. Datta, Electronic Transport in Mesoscopic Systems. Cambridge University Press, 1995. \cite{Datta1995}

[30] S. Datta, Quantum Transport: from Atom to Transistor. Cambridge University Press, 2005. \cite{Datta2005}

[31] T. Fang, A. Konar, H. Xing, and D. Jena, “Carrier Statistics and Quantum Capacitance of Graphene Sheets and Ribbons,” Applied Physics Letters, vol. 91, no. 9, p. 092109, 2007. \cite{Fang2007}

[32] Y.-W. Tan et al., “Measurement of Scattering Rate and Minimum Conductivity in Graphene,” Physical Review Letters, vol. 99, no. 24, p. 246803, Dec. 2007. \cite{Tan2007}

[33] C. Canali, G. Majni, R. Minder, and G. Ottaviani, “Electron and hole drift velocity measurements in silicon and their empirical relation to electric field and temperature,” IEEE Transactions on Electron Devices, vol. 22, no. 11, pp. 1045-1047, Nov. 1975. \cite{Canali1975}

[34] A. Khakifirooz, O. M. Nayfeh, and D. Antoniadis, “A Simple Semiempirical Short-Channel MOSFET Current–Voltage Model Continuous Across All Regions of Operation and Employing Only Physical Parameters,” IEEE Transactions on Electron Devices, vol. 56, no. 8, pp. 1674-1680, Aug. 2009. \cite{Khakifirooz2009}

## Figures

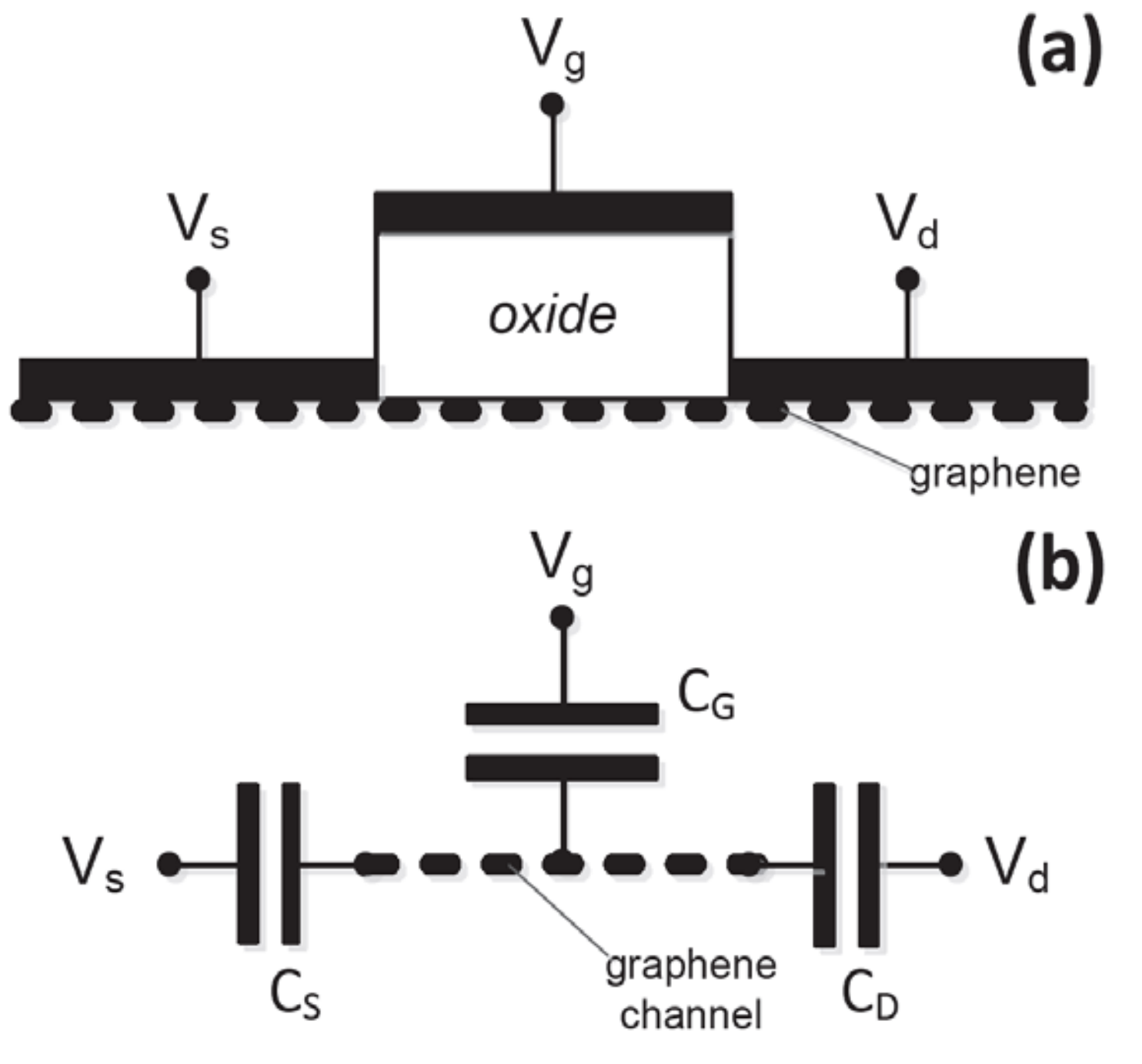

Fig. 1. (a) The cross-section of a graphene FET. Graphene is the black thick dotted line under contacts and gate oxide. (b) The lumped element model for the electrostatic part of the problem.

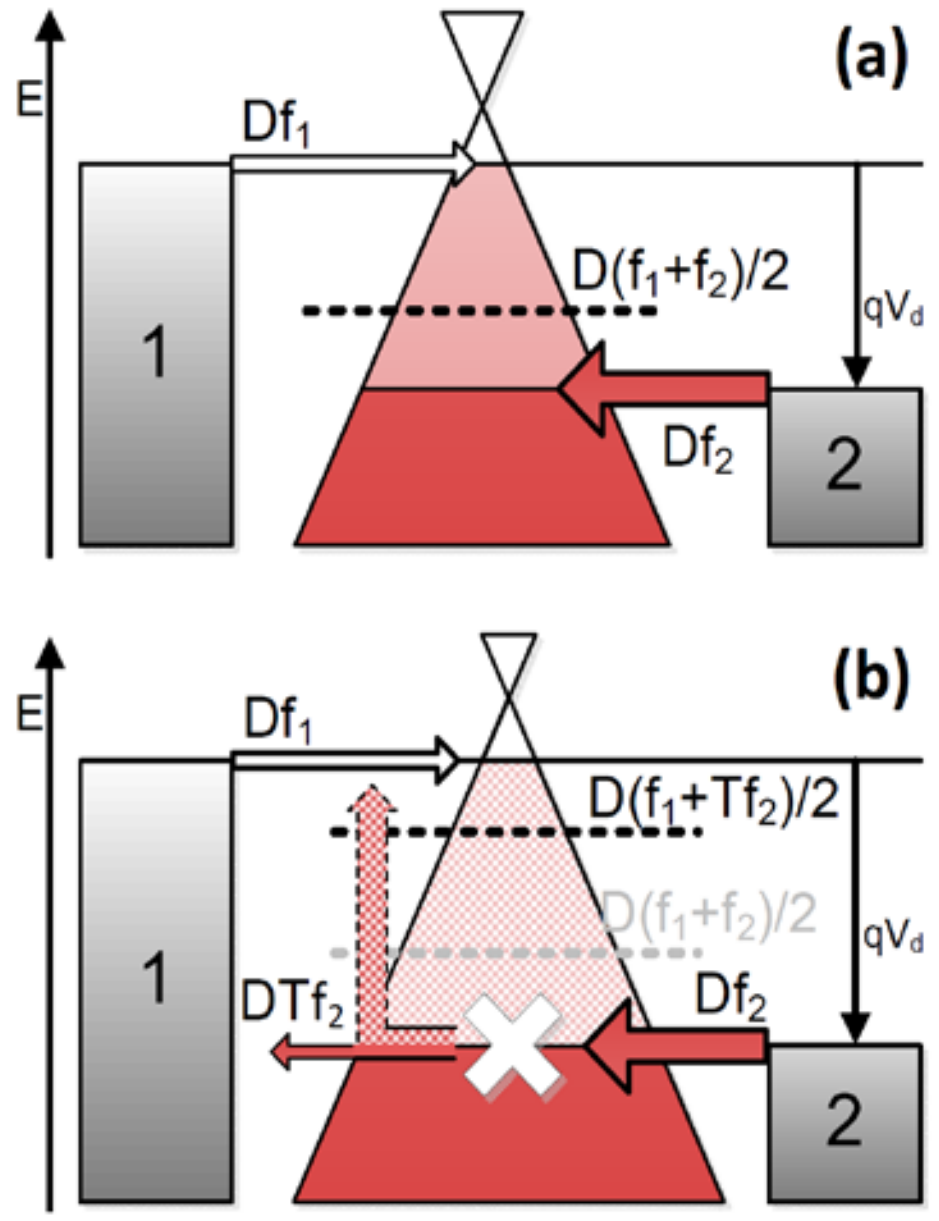

**Figure 2.** (a) Injection of carriers into the channel. The double cone is the density of states of large-area graphene, while "1" and "2" respectively are the pseudo-Fermi levels of source and drain contacts. The black dashed line between the two injected fluxes represents the population of carriers, according to the two contacts. (b) Same context of before but including an isolated scatterer (the big white cross). A fraction $(1 - T)$ of the drain flux, going upwards with dashed contour, is scattered and reaches thermal equilibrium. The black dashed line represents the population of carriers, according to the effective pseudo-Fermi levels of contacts.

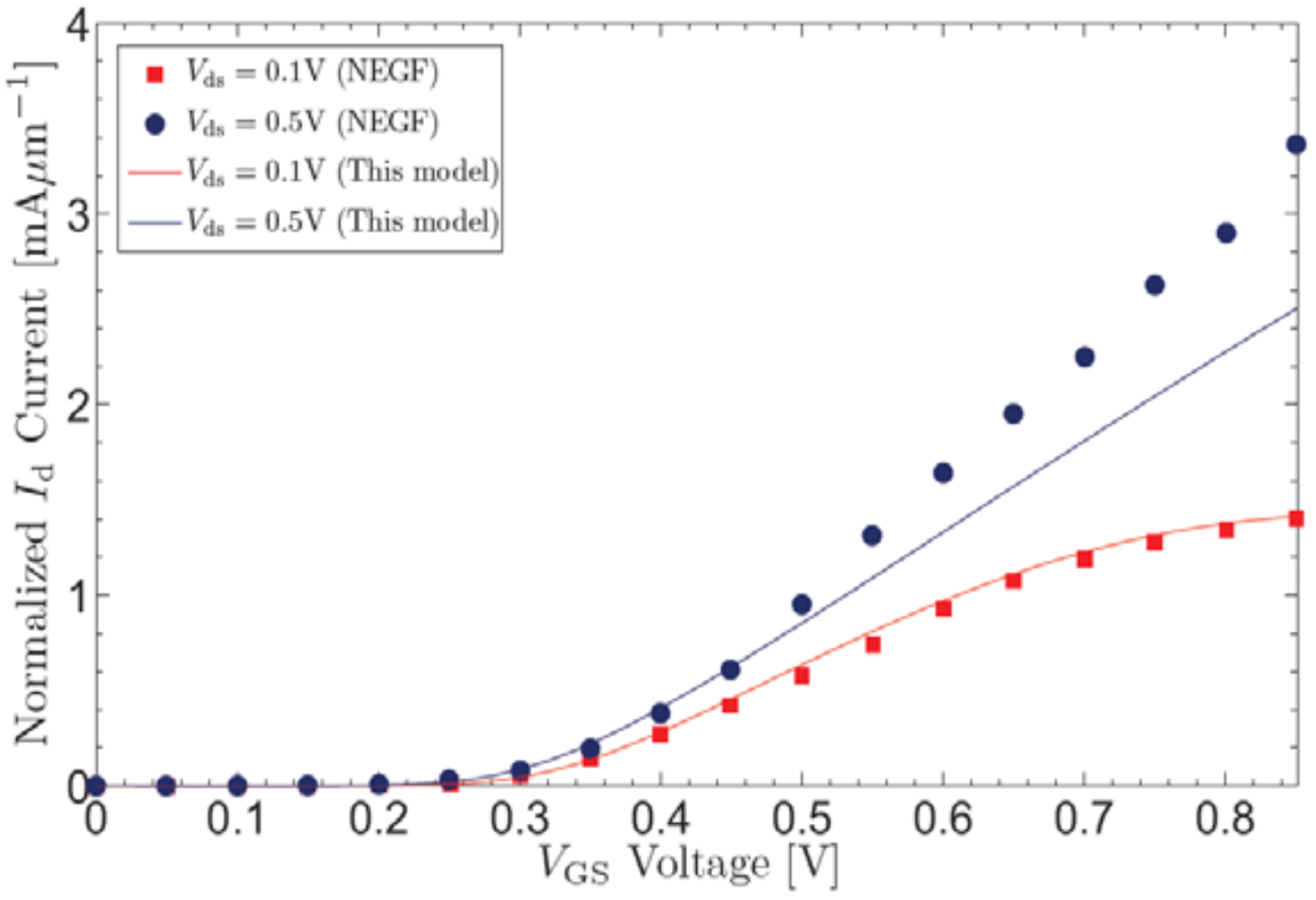

Fig. 3. Transfer Characteristic in linear scale for this model compared to NEGF from [12].

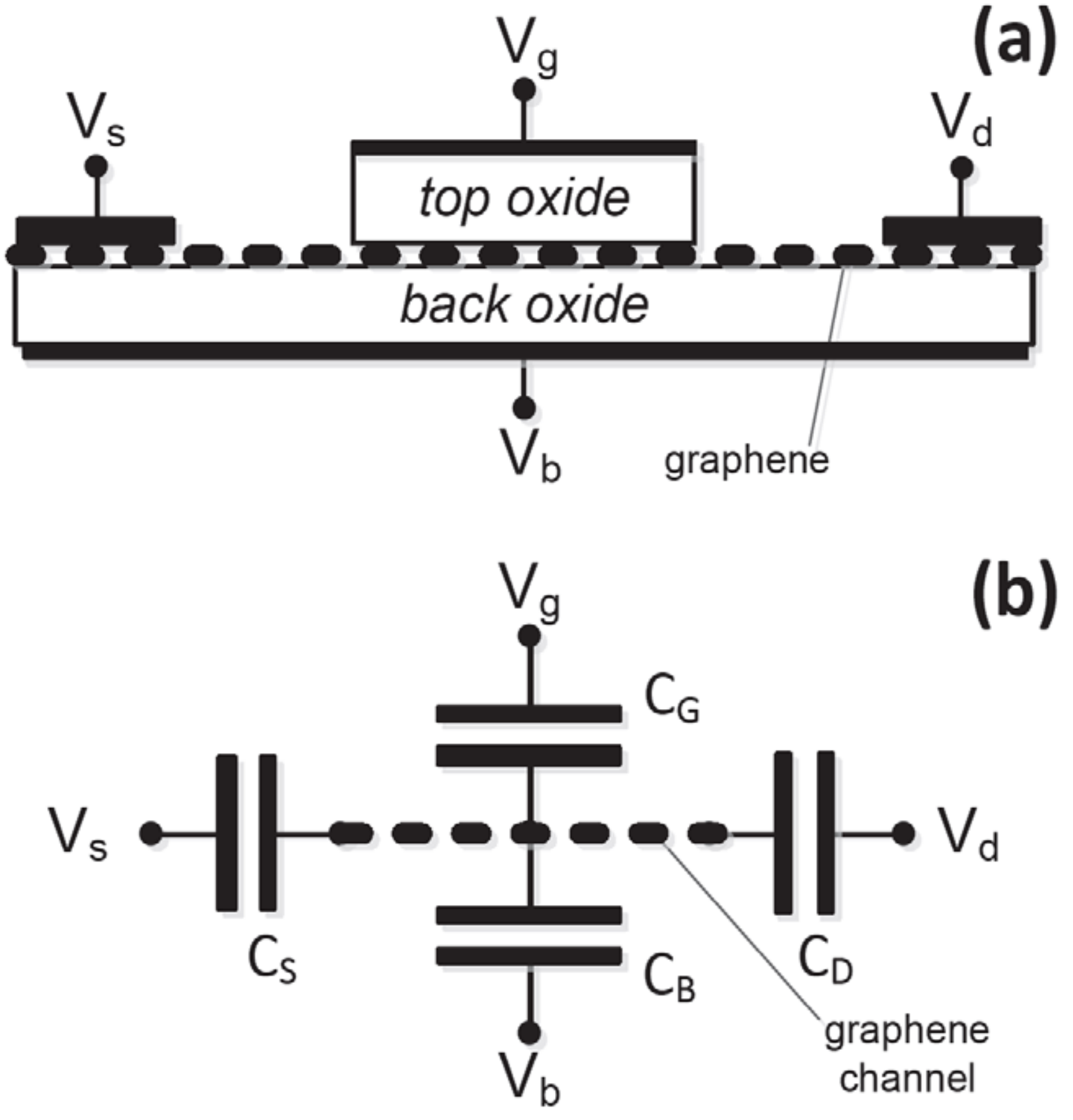

Fig. 4. (a) The cross-section of the large-area graphene FET of ref [20]. Graphene is the black thick dotted line under contacts and top-gate oxide. A global back-gate electrode is shown under the back-gate oxide. (b) The lumped element model that accounts for the electrostatics of a large-area double-gate device.

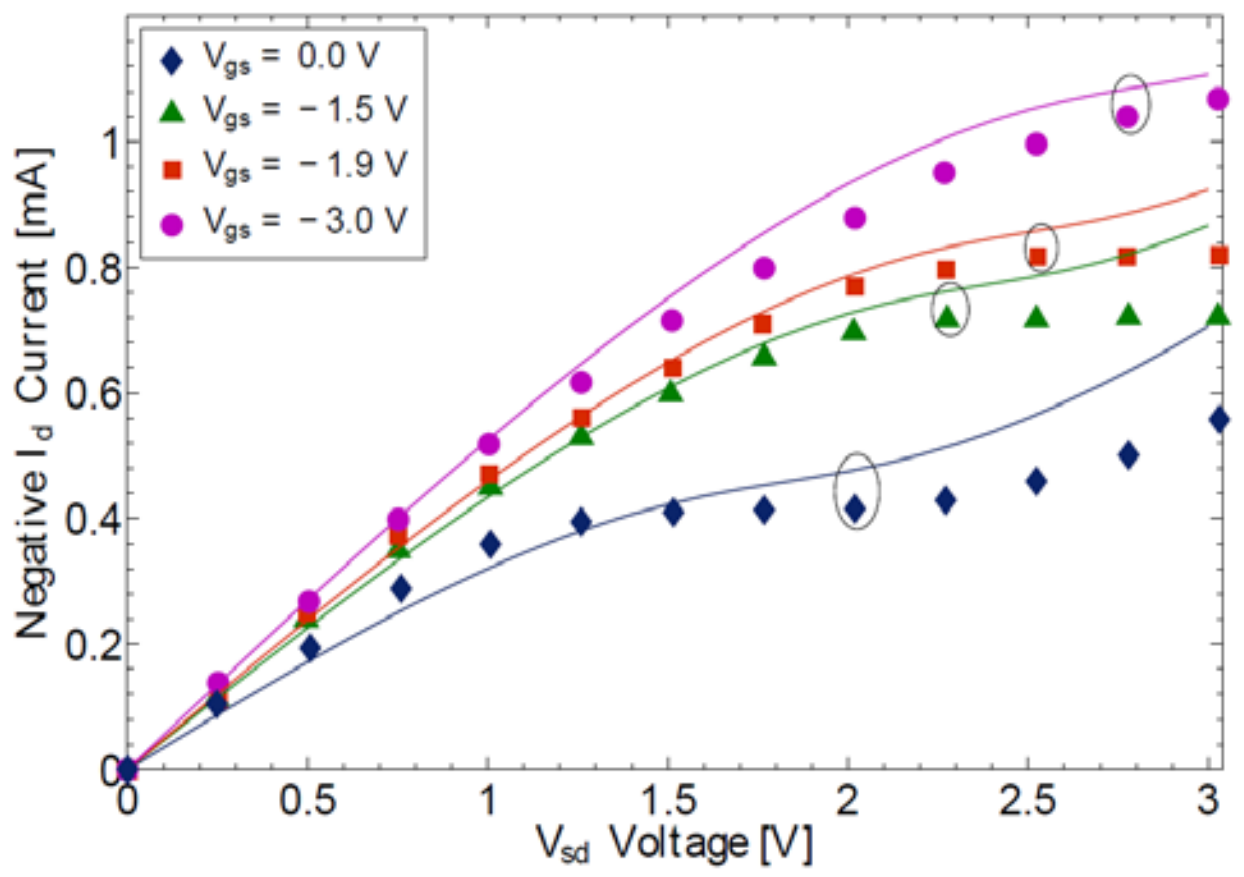

**Figure 5.** Simulated $I_\mathrm{d}\,(V_\mathrm{sd})$ characteristic for FET2 (solid lines) for Gate voltages from $V_\mathrm{GS-top} = 0\mathrm{V}$ to $-3\mathrm{V}$ compared to the experimental values (markers only) of the current.